\documentclass[10pt,conference]{IEEEtran}

\usepackage{microtype}
\usepackage{graphicx}
\usepackage{subcaption}
\usepackage{booktabs}
\usepackage{amsmath}
\usepackage{amssymb}
\usepackage{mathtools}
\usepackage{amsthm}
\usepackage{url}
\usepackage{float}
\usepackage{wrapfig}
\usepackage{siunitx}
\usepackage{makecell}
\usepackage{svg}
\usepackage{orcidlink}
\usepackage{tikz}
\usetikzlibrary{quantikz2}

\usepackage{hyperref}
\usepackage[capitalize,noabbrev]{cleveref}
\crefname{figure}{Figure}{Figures}
\Crefname{figure}{Figure}{Figures}
\crefname{table}{Table}{Tables}
\Crefname{table}{Table}{Tables}
\crefname{section}{Section}{Sections}
\Crefname{section}{Section}{Sections}
\crefname{appendix}{Appendix}{Appendices}
\Crefname{appendix}{Appendix}{Appendices}
\crefname{equation}{Equation}{Equations}
\Crefname{equation}{Equation}{Equations}

\usepackage[textsize=tiny]{todonotes}

\theoremstyle{plain}

\theoremstyle{definition}

\theoremstyle{remark}

\begin{document}

\title{Mitigating Exponential Mixed Frequency Growth in QML}

\author{
  \IEEEauthorblockN{Michael Poppel\orcidlink{0009-0005-1141-0974}\IEEEauthorrefmark{1}\IEEEauthorrefmark{2},
                    David Bucher\orcidlink{0009-0002-0764-9606}\IEEEauthorrefmark{2},
                    Maximilian Zorn\orcidlink{0009-0006-2750-7495}\IEEEauthorrefmark{1},
                    Nico Kraus\orcidlink{0009-0003-2329-6166}\IEEEauthorrefmark{2},
                    Claudia Linnhoff-Popien\orcidlink{0000-0001-6284-9286}\IEEEauthorrefmark{1},\\
                    Philipp Altmann\orcidlink{0000-0003-1134-176X}\IEEEauthorrefmark{1},
                    Jonas Stein\orcidlink{0000-0001-5727-9151}\IEEEauthorrefmark{1}}\\
  \IEEEauthorblockA{\IEEEauthorrefmark{1}Department of Computer Science, LMU Munich, Germany}
  \IEEEauthorblockA{\IEEEauthorrefmark{2}Aqarios GmbH, Munich, Germany}
}

\maketitle

\begin{abstract}
Angle encoding has emerged as a popular feature map for embedding 
classical data into quantum models, naturally generating truncated 
Fourier series with universal function approximation capabilities. 
Despite this expressive capability, practical training faces 
significant challenges. Through controlled experiments with white-box 
target functions, we demonstrate that training failures can occur 
even when all established parameter sufficiency conditions are 
satisfied. Building on the redundancy-gradient framework of Duffy 
and Jastrzebski, we provide systematic experimental evidence that 
non-unique frequencies dominate the gradient landscape and crowd out 
target frequencies --- a burden that grows exponentially with 
encoding depth under unary encoding. Small-angle initialization mitigates this in one-dimensional 
settings but fails to scale to higher dimensions, where even 
ternary encoding --- which minimizes per-frequency redundancy --- 
faces intractable combinatorial growth of unique frequency tuples 
regardless of initialization or optimizer choice. We introduce frequency selection as a principled 
solution that restricts the model spectrum to only those frequencies 
present in the target function. For two-dimensional targets, 
frequency selection achieves near-optimal performance (median 
$R^2 \approx 0.95$) where dense approaches struggle, and remains 
tractable at high-frequency magnitudes where dense approaches fail 
entirely (median $R^2 \approx 0.85$). Validation on a real-world 
dataset confirms the approach transfers beyond synthetic settings.
\end{abstract}


\section{Introduction}

The identification of latent relationships within complex, 
high-dimensional datasets represents a fundamental challenge across 
numerous scientific and commercial domains. Drug development, for 
example, requires modeling quantum effects such as electron 
delocalization and chemical bonding, which leads to exponential 
scaling when computed classically \cite{ijms26136325}. Financial 
markets exhibit complex feature dependencies that traditional modeling 
approaches struggle to capture, as they typically assume constant 
correlations and neglect tail dependencies \cite{EasyChair:15051}.

Quantum machine learning (QML) emerges as a promising paradigm to 
address these challenges, offering theoretical advantages over 
classical approaches. Studies demonstrate that quantum models can 
represent more complex functions than classical counterparts 
\cite{Mitarai_2018}, exhibiting higher effective dimensionality and 
faster training compared to traditional neural networks \cite{Abbas_2021}.

Substantial challenges remain for QML, however. Barren plateaus limit 
the number of trainable qubits \cite{Ragone_2024}, while NISQ hardware 
constraints hinder practical deployment. Although qubit counts continue 
rising and error rates are improving, circuit depth must remain shallow 
to mitigate noise accumulation, making efficient parameter utilization 
critical.

Analogous to activation functions in classical neural networks, quantum 
circuits achieve non-linearity through angle encoding, data re-uploading, 
and entangling gates---mechanisms that, as shown by \cite{Schuld_2021}, 
enable quantum models to represent truncated Fourier series. This Fourier 
structure carries a deeper significance: Belis et al.\ \cite{Belis_2026} 
argue that spectral methods are fundamental to machine learning, and that 
angle encoding naturally induces a simplicity bias analogous to the 
spectral bias observed in deep learning \cite{Rahaman_2019}, which 
preferentially learns lower-order Fourier components before higher-order 
ones. Angle encoding models therefore inherit a form of implicit 
regularization through their restricted frequency spectrum---a property 
that, as we show, also constrains what is achievable without careful 
frequency management. For successful training in supervised learning 
tasks, the following conditions have been established as necessary: 
1) having a sufficient model frequency spectrum, together with 2) at 
least an equal number of parameters to control the model frequencies, 
and 3) enough parameters to explore all independent directions within the orbit of reachable states \cite{Schuld_2021, Larocca_2023}.

Through white-box experiments with known target functions, we demonstrate 
that these conditions are not sufficient: the model also contains 
non-unique frequencies whose redundancy $R(\omega)$, as formalized by 
Duffy and Jastrzebski~\cite{Duffy_2026}, drives a spectral bias that 
systematically crowds out higher target frequencies during training. With exponential growth of total redundancy under unary encoding, 
and combinatorial growth of unique frequency tuples under ternary 
encoding in higher dimensions, this burden can be mitigated to a 
limited extent by small-angle initialization of the weight 
parameters, but cannot be overcome by initialization or optimizer 
choice alone. It is therefore essential to reduce the number 
of model frequencies to those present in the target function. We propose 
including only selected frequencies in the model spectrum rather than a 
dense spectrum containing numerous unnecessary elements, achieving 
consistently superior $R^2$ scores with lower variance than the standard 
dense frequency approach.

We summarize our contributions as follows:
\begin{itemize}
    \item A systematic analysis consolidating existing parameter sufficiency 
    conditions, revealing that non-unique frequencies impose an 
    additional training burden not fully captured by existing theory. 
    Under unary encoding this burden grows exponentially with encoding 
    depth due to high redundancy; more broadly, for both encodings, 
    the combinatorial growth of unique frequency tuples with input 
    dimensionality and target frequency magnitude renders dense 
    approaches intractable as problem scale increases.
    \item An empirical demonstration that small-angle initialization 
    according to Zhang et al.\ \cite{Zhang_2025} mitigates 
    redundancy-driven spectral bias in one-dimensional settings, 
    but that as dimensionality increases the dominant failure mode 
    shifts to resource intractability: satisfying the frequency 
    control condition $p \geq |\Omega|$ for dense encodings demands 
    a parameter budget that grows prohibitively with target frequency 
    magnitude and input dimension, which no initialization strategy 
    can resolve.
    \item A frequency selection approach that addresses the redundancy 
    burden at its source, enabling more effective use of the available 
    parameter budget and achieving competitive performance where dense 
    frequency strategies fail.
\end{itemize}

\section{Theoretical Background}
\label{sec:background}

\subsection{Variational Quantum Circuits}

Variational quantum circuits (VQCs) represent a prominent class of QML 
algorithms that integrate quantum operations with classical optimization. 
Classical data is encoded into a quantum state $|\Psi_0(x)\rangle = 
S(x)\ket{0}$ via a feature map $S(x)$, then processed by a parametrized 
ansatz $W(\boldsymbol{\theta})$, yielding 
$|\Psi(x, \boldsymbol{\theta})\rangle = W(\boldsymbol{\theta})S(x)\ket{0}$. 
Measuring an observable $M$ produces the model output:
\begin{equation}
    f(x, \boldsymbol{\theta}) = \langle\Psi(x, \boldsymbol{\theta})|
    M|\Psi(x, \boldsymbol{\theta})\rangle
\end{equation}
For a regression task with training set $\{(x_i, y_i)\}_{i=1}^N$, 
the loss function is the mean squared error between model predictions 
and target values:
\begin{equation}
    \mathcal{L}(\boldsymbol{\theta}) = \frac{1}{N}\sum_{i=1}^N 
    \left(f(x_i, \boldsymbol{\theta}) - y_i\right)^2
\end{equation}
A classical optimizer updates the parameters iteratively to minimize 
$\mathcal{L}$:
\begin{equation}
    \boldsymbol{\theta}^* = \underset{\boldsymbol{\theta}}{\mathrm{argmin}} 
    \,\mathcal{L}(\boldsymbol{\theta})
\end{equation}
This quantum-classical loop continues until convergence~\cite{farhi2018classificationquantumneuralnetworks}.

\subsection{Quantum Models as Fourier Series}
\label{sec:quantum_fourier_series}
Multiple FMs and ansatz layers are combined into a parameterized 
quantum circuit $U(x, \boldsymbol{\theta})$, alternating $L$ angle 
encoding FMs $S(x)$ with $L+1$ trainable ansatz layers:
\begin{equation}
    U(x, \boldsymbol{\theta}) = W^{(L)}(\boldsymbol{\theta_{L+1}})
    S(x)W^{(L)}(\boldsymbol{\theta_L}) \ldots 
    S(x)W^{(0)}(\boldsymbol{\theta_1})
    \label{eq:pqc}
\end{equation}
As shown by \cite{Schuld_2021}, the model output naturally represents 
a partial Fourier series:
\begin{equation}
    f_{\boldsymbol{\theta}}(x) = \bra{0} U^{\dagger}(x, 
    \boldsymbol{\theta}) MU(x, \boldsymbol{\theta})\ket{0}
    = \sum_{\omega \in \Omega} c_{\omega}e^{i\omega x}
    \label{eq:fourier_series}
\end{equation}
For angle encoding FMs $S(x) = e^{-i\frac{x}{2}\sigma}$ with Pauli 
operator $\sigma$, the frequency spectrum $\Omega$ is determined by 
the eigenvalue structure of the encoding Hamiltonian. For the parallel 
architecture distributing $L$ FMs across $L$ qubits, the joint 
encoding unitary diagonalizes as $S_p(x) = Ve^{-ix\Sigma}V^\dagger$, 
where $\Sigma = \mathrm{diag}(\lambda_1, \ldots, \lambda_{2^L})$ 
contains the joint eigenvalues \cite{Schuld_2021}. The spectrum $\Omega$ comprises all 
pairwise eigenvalue differences $\Lambda_k - \Lambda_j$, yielding 
$(2^L)^2 = 4^L$ total frequencies of which only $2L+1$ are 
unique \cite{Schuld_2021}. The remaining frequencies are non-unique, 
arising because different pairs $(k,j)$ produce identical differences. 
Following \cite{Duffy_2026}, we define the \emph{redundancy} $R(\omega)$ 
as the number of such contributing pairs, with $\omega \in \{-L, \ldots, L\}$, 
which evaluates to a binomial coefficient \cite{Peters_2023, holzer2024spectralinvariancemaximalityproperties}:
\begin{equation}
    R(\omega) = \left|\left\{(k,j) \mid \Lambda_k - \Lambda_j = 
    \omega\right\}\right| = \binom{2L}{L - |\omega|}
    \label{eq:redundancy}
\end{equation}
This yields a symmetric, bell-shaped redundancy profile peaking at 
$R(0) = \binom{2L}{L}$ and falling to $R(\pm L) = 1$ at the extremes. 
As shown in \cite{Duffy_2026}, this induces a natural spectral bias: 
since gradient magnitudes are bounded by $R(\omega)$, the optimizer 
preferentially learns low-frequency components, which receive 
disproportionately large gradient signals compared to higher frequencies.
The Fourier coefficients decompose accordingly as:
\begin{equation}
    c_\omega = \sum_{\substack{k,j \in [d]^L \\ 
    \Lambda_k - \Lambda_j = \omega}} a_{k,j}
    \label{eq:coeff_decomp}
\end{equation}
The total non-unique frequency count $\sum_{\omega}(R(\omega)-1) = 
4^L - (2L+1)$ grows exponentially with encoding layers while the 
unique count grows only linearly. The serial 
architecture places $L$ FMs sequentially on a single qubit and 
generates the same spectrum and redundancies \cite{holzer2024spectralinvariancemaximalityproperties}.

\subsection{Encoding Strategies and Frequency Spectrum Generation}
The frequency spectrum can be expanded by multiplying individual FM 
Hamiltonians $\sigma_i$ by prefactors $p_i$, which scales the 
corresponding eigenvalues and reduces redundancies arising from 
degenerate eigenvalues and identical differences, thereby reducing 
$R(\omega)$ for each $\omega \in \Omega$.

Ternary encoding employs prefactors $p_i = 3^i$ with 
$i \in \{0, \ldots, L-1\}$, generating a dense spectrum of $3^L$ 
unique frequencies---the largest possible cardinality for separable 
encoding generators \cite{Peters_2023, Kordzanganeh_2023, Shin_2023}:
\begin{equation}
    |\Omega| = 3^L = \left| \left\{ -\left\lfloor \frac{3^L}{2} 
    \right\rfloor, \ldots, 0, \ldots,  \left\lfloor 
    \frac{3^L}{2} \right\rfloor \right\} \right|
    \label{eq:spectrum_cardinality}
\end{equation}
In the following, we focus on \emph{unary encoding} (prefactors $p_i = 1$) 
and \emph{ternary encoding} (prefactors $p_i = 3^i$) as the two 
extremes of this redundancy--expressivity trade-off.

\section{Parameter Requirements for Frequency Control and Scaling 
Challenges}
\label{sec:param_freq}

A fundamental constraint in quantum machine learning emerges from the 
relationship between frequency control and parameter requirements. For 
a quantum model to fully utilize its frequency spectrum, each unique 
frequency must possess independent control through at least one dedicated 
parameter \cite{Schuld_2021}:
\begin{equation}
    p \geq |\Omega| \quad \textnormal{for} \; \boldsymbol{\theta} \in 
    \mathbb{R}^p
    \label{eq:parameters}
\end{equation}
As will be seen in \cref{sec:results}, this is a necessary but not 
sufficient requirement. Crucially, the parameters counted here must be 
\emph{individually trainable}. A single qubit provides only 
2 linearly independent directions in its state space regardless of 
circuit depth \cite{Larocca_2023}, and consecutive rotations on the 
same qubit axis collapse into a single effective parameter 
\cite{nielsen2002quantum}. Only when non-linearity through feature 
maps or entangling gates separates parameter rotations can additional 
parameters be trained independently. Without sufficient individually 
trainable parameters, the optimizer cannot exert independent control 
over each frequency component in $\Omega$, regardless of how many 
parameters are nominally present in the circuit.

\subsection{Multi-dimensional Extensions and Mixed Frequencies}

For datasets containing $d$ features, interactions between feature 
encodings generate mixed frequencies, and the overall frequency spectrum 
becomes the Cartesian product of the individual spectra 
\cite{holzer2024spectralinvariancemaximalityproperties}:
\begin{equation}
    \Omega = \Omega_1 \times \ldots \times \Omega_d, \quad
    |\Omega| = \prod_{i=1}^d|\Omega_i|
    \label{eq:multiDim_freq_cardinality}
\end{equation}
With $r$ repeated ternary feature maps per dimension, $|\Omega_i| = 3^r$ 
and the total spectrum cardinality scales as $(3^r)^d$ — exponential 
in both encoding depth and input dimension. The parameter requirement 
from \cref{eq:parameters} scales accordingly, rendering independent 
control of all unique frequencies intractable as dimensionality grows. 
Crucially, the total redundancy burden $\sum_\omega R(\omega) = (4^r)^d$ 
grows even faster, compounding the optimization difficulty through 
non-unique frequency dominance of the gradient landscape.

\subsection{Avoiding Spurious Local Minima with Ansatz Overparameterization}

The $L+1$ ansatz layers $W^{(i)}$ in \cref{eq:pqc} are typically 
constructed from one or more identical blocks $W^{(i)}_j$ to facilitate 
scalability, as detailed in \cref{sec:circuit_architectures}. Each 
block operates on multiple qubits, but is composed of elementary gates 
acting on one- or two-qubit subsystems: $W^{(i)}_j = \prod_k 
W^{(i)}_{j,k} = \prod_k e^{a_k}$, where the set of generators is 
defined as $\mathcal{G} = \{a_k\}_{k=1}^K$ and $K$ represents the 
total number of generators within an ansatz block \cite{Wiersema_2024}. 
A widely adopted ansatz architecture is the Hardware Efficient Ansatz 
(HEA) \cite{Kandala_2017}, which combines parameterized rotational gates 
with entanglement gates such as CNOTs. In this framework, the ansatz 
generators $\{a_k\}$ consist of single-qubit Pauli rotations $\sigma 
\in \{\sigma_x, \sigma_y, \sigma_z\}$ or the identity $I$, and tensor 
products thereof for two-qubit operations, such as $\pi/4 (Z \otimes I 
- Z \otimes X)$ for the CNOT gate. Since combinations of generators can 
produce additional generators through commutation relations, it becomes 
essential to characterize the DLA $\mathfrak{g}$, which describes the 
complete set of unitary evolutions accessible to a given generator set. 
The DLA is defined through the Lie closure of the generators:
\begin{equation}
    \mathfrak{g} = \textnormal{span} \langle ia_1, \ldots,
    ia_K \rangle_{\mathrm{Lie}}
\end{equation}
where the Lie closure is formed by repeated nested commutators of the 
generators in $\mathcal{G}$. The DLA fundamentally determines the set 
of reachable unitaries $U(x, \boldsymbol{\theta})$ and thereby the 
orbit of reachable pure states $\mathcal{O} = \{U|\psi\rangle : U \in 
e^{\mathfrak{g}}\}$ accessible from the encoded input states 
$\Psi_0(x)$.

Work from \cite{Larocca_2023} reveals that spurious local minima 
can be avoided when the number of ansatz parameters is sufficient 
for the model to explore all independent directions within the orbit 
$\mathcal{O}$. The orbit dimension is bounded both by 
$\textnormal{dim}(\mathfrak{g})$ and by the dimension of the pure 
state manifold $\mathbb{CP}^{d-1}$:
\begin{equation}
    \dim(\mathcal{O}) \leq 
    \min\!\left(\textnormal{dim}(\mathfrak{g}),\ 2(d-1)\right)
\end{equation}
where $d = 2^n$ is the Hilbert space dimension. For fully expressive ansätze such as the HEA with universal gate 
sets, $\textnormal{dim}(\mathfrak{g}) = 4^n - 1 \gg 2(d-1) = 
2^{n+1}-2$, so the pure state manifold dimension is the tighter 
constraint (for ansätze with smaller DLAs, such as problem-inspired 
ansätze, $\textnormal{dim}(\mathfrak{g})$ may itself be the binding 
bound). The sufficient parameter condition for overparameterization 
therefore becomes:
\begin{equation}
    p \geq 2^{n+1} - 2
    \label{eq:op_condition}
\end{equation}
irrespective of the encoding strategy employed.

\subsection{Weight Initialization}
\label{sec:initialization}
As established in \cref{sec:quantum_fourier_series}, the redundancy 
profile $R(\omega)$ creates a spectral bias toward low-frequency 
components: gradient magnitudes are bounded above by $R(\omega)$ 
\cite{Duffy_2026}, so non-unique frequencies attract disproportionately 
large gradient signals and are fitted preferentially during training, 
while target frequencies receive smaller gradient signals and are 
systematically deprioritized.

\emph{Small-angle initialization} \cite{Zhang_2025, Duffy_2026} 
addresses this by suppressing all initial Fourier coefficients toward 
zero, reducing the redundancy-driven gradient dominance and giving 
target frequencies a fairer share of the gradient signal from the 
start of training. Specifically, Zhang et al.\ \cite{Zhang_2025} 
propose initializing parameters from a Gaussian distribution
\begin{equation}
    \theta \sim \mathcal{N}\!\left(0,\, \frac{1}{L}\right)
    \label{eq:zhang_init}
\end{equation}
to avoid exponentially vanishing gradients. 

\subsection{Summary of Requirements}
\label{sec:param_requirements}
The analysis presented in the preceding sections reveals three 
fundamental conditions that must be satisfied for effective quantum 
machine learning model training:
\begin{enumerate}
    \item \textbf{Frequency coverage condition}: The frequencies 
    $\omega_t$ present in the target function must be contained within 
    the model's frequency spectrum $\Omega_m$: $\omega_t \in \Omega_m$. 
    This ensures that the quantum model possesses the necessary spectral 
    components to represent the target function.
    \item \textbf{Frequency control condition}: For complete utilization 
    of all frequencies in the model spectrum, the number $p_{it}$ of 
    individually trainable ansatz parameters must equal or exceed the 
    cardinality of the frequency spectrum: $p_{it} \geq |\Omega|$. This 
    condition allows individual control over each unique frequency 
    component.
    \item \textbf{Optimization landscape condition}: Spurious local 
    minima can be avoided when the number of individually trainable 
    parameters is sufficient for the model to explore all independent 
    directions within the orbit of reachable states. For fully 
    expressive ansätze this requires $p_{it} \geq 2^{n+1} - 2$ 
    (see \cref{eq:op_condition}), while for ansätze with smaller 
    DLAs the binding bound is $p_{it} \geq \textnormal{dim}(\mathfrak{g})$ 
    \cite{Larocca_2023}.
\end{enumerate}
These three conditions collectively establish the minimum parameter 
requirements for quantum machine learning models, with the effective 
parameter count being determined by $p_{it} = \max\{|\Omega|, 
2^{n+1}-2\}$ for fully expressive ansätze, subject to the 
constraint that the target function frequencies are representable 
within the chosen encoding scheme.

\section{Frequency Selection and Near-Zero Weight Initialization}
\label{sec:methodology}

\subsection{Circuit Architectures and Experimental Design}
To investigate parameter sufficiency requirements, we employ 
one-dimensional white-box target functions that allow precise control 
over theoretical parameter needs. We examine four circuit architectures 
that together probe each of the conditions established in 
\cref{sec:param_requirements}: single-qubit serial encoding with unary 
and ternary prefactors, and multi-qubit parallel encoding with unary 
and ternary prefactors. Example circuit diagrams are provided in 
\cref{sec:circuit_architectures}.
As established in \cref{sec:param_freq}, individually trainable 
parameters are distinct from linearly independent ones. A single qubit 
has DLA dimension 3, yet Perez-Salinas et al.\ \cite{Perez_Salinas_2021} 
demonstrate that a single-qubit re-uploading model is a universal 
function approximator. In our serial architectures, feature maps between 
ansatz layers prevent parameter collapse, so the individually trainable 
parameter count grows with circuit depth even though the DLA dimension 
remains fixed at 3. Any fitting failure in the serial unary setting 
therefore cannot be attributed to the DLA constraint.

Instead, we investigate whether the spectral bias induced by frequency 
redundancy \cite{Duffy_2026} acts as a double-edged sword: while it 
provides implicit regularization by suppressing high-frequency noise 
during early training, it simultaneously prevents gradient-based 
optimizers from fitting higher target frequencies, whose gradient 
signals are systematically dominated by low-frequency, high-redundancy 
components. By contrasting unary and ternary encoding schemes---which 
generate substantially different redundancy profiles $R(\omega)$ for 
comparable frequency coverage---we study this spectral bias empirically 
in \cref{sec:results}.

To investigate whether the influence of frequency redundancies on 
gradient-based optimization can be overcome, we additionally evaluate 
gradient-free optimizers (CMA-ES, Nelder-Mead, and SPSA), 
which bypass the gradient magnitude bounds established in 
\cite{Duffy_2026}.

\subsection{Small-Angle Weight Initialization}

As established in \cref{sec:initialization}, small-angle initialization 
reduces the redundancy-driven gradient dominance by suppressing all 
Fourier coefficients toward zero at the start of training. In our 
experiments we employ the adaptive Gaussian 
scheme of Zhang et al.\ \cite{Zhang_2025} as defined in 
\cref{eq:zhang_init}.

\subsection{Frequency Selection}
\label{sec:freq_selection}

Quantum models employing ternary encoding generate the largest dense 
frequency spectrum achievable for separable encoding generators. While 
this comprehensive spectral coverage ensures that potential target 
function frequencies are available, it necessitates exponential 
parameter scaling that often renders practical implementation 
infeasible. However, when the target function's spectral composition 
is known a priori, this complete frequency coverage becomes unnecessary 
and computationally wasteful, as the optimizer must contend with the 
full redundancy burden of the dense spectrum rather than only those 
frequencies relevant to the target.

By selecting prefactors that deviate from the standard ternary 
sequence---either through manual selection, systematic trial-and-error 
approaches, or dedicated classical optimization---we can exploit the 
eigenvalue structure of the encoding Hamiltonians to construct sparse 
frequency spectra that ideally encompass only the frequencies present 
in the target function, directly reducing $|\Omega|$ and with it both 
the parameter requirement and the optimization burden.

To illustrate the reduction in spectral complexity achieved by frequency 
selection, consider a two-feature-map configuration with $R_x$ encoding 
gates and prefactors $\{3, 9\}$. The eigenvalue differences 
yield the sparse frequency spectrum $\Omega = \{-12, -9, -6, -3, 0, 3, 
6, 9, 12\}$ containing only 9 frequencies. In contrast, the equivalent 
dense ternary spectrum with prefactors $\{1, 3, 9\}$ that would 
naturally contain the target frequencies $\{3, 6, 9, 12\}$ spans 
$\{-13, -12, \ldots, 0, \ldots, 12, 13\}$ with 27 frequencies. 
The selected spectrum therefore reduces the optimization burden from 
$4^3 = 64$ to $4^2 = 16$ non-unique frequency instances, while still 
covering all target frequencies.

\section{Empirical Evaluation}
\label{sec:results}

\subsection{Experimental Setup}
\label{sec:experimental_setup}

Our experimental approach proceeds in two stages. First, we evaluate 
several circuit architectures using a 1D target function to identify 
which configurations can satisfy the parameter requirements established 
in \cref{sec:param_requirements}. Subsequently, we extend the 
investigation to two-dimensional target functions using the 
architectures that successfully meet these requirements, comparing 
dense frequency approaches under two initialization strategies 
against frequency selection.

We employ white-box target functions in the form of real-valued partial 
Fourier series $t\colon [-\pi, \pi]^d \to \mathbb{R}$ with target 
frequencies $\{3, 6, 9, 12\}$ for the 1D case, $\{3, 6, 9, 12\}^2$ 
for the low-frequency 2D case, and $\{10, 20, 30, 40\}^2$ for a 
high-frequency 2D case that further stresses the redundancy competition 
condition. For the 2D experiments we compare dense ternary encoding under uniform 
and Zhang initialization against selected frequency encoding with 
prefactors chosen to cover only the target spectrum. For the 
low-frequency case, the dense model requires $|\Omega| = 3^4 = 81$ 
unique frequencies per dimension and $81^2 = \num{6561}$ frequency pairs in total, 
which already strains the parameter budget. For the high-frequency 
case, covering frequencies up to 40 with ternary prefactors 
$\{1,3,9,27\}$ yields $|\Omega| = 81^2 = \num{6561}$ unique frequency 
pairs --- far exceeding the 480-parameter budget and therefore 
violating condition 2 by design. Frequency selection 
with prefactors $\{10,30\}$ reduces this to $9^2 = 81$ unique 
frequency pairs within the same budget.

In addition to the synthetic experiments, we evaluate frequency 
selection on the Combined Cycle Power Plant (CCPP) 
dataset~\cite{combined_cycle_power_plant_294} as a real-world 
validation using a hardware-native ladder ansatz. We compare full 
ternary encoding against a selected-frequency model on a four-qubit 
parallel architecture, and additionally assess robustness under 
depolarizing noise. Full details are provided in 
\cref{sec:ccpp_setup}.

\subsection{Architecture Selection and Spectral Bias}
\label{sec:param_requirements_results}

\subsubsection{1D Synthetic Target Functions}
\label{sec:1d_param_requirements}

We begin our investigation with 1D partial Fourier series using model 
configurations that provide dense frequency spectrum coverage of all 
target frequencies, in order to identify suitable circuit architectures 
and to avoid fitting failure due to unsuitable experimental set-up.

\begin{figure}[htbp]
    \centering
    \includegraphics[width=0.48\textwidth]{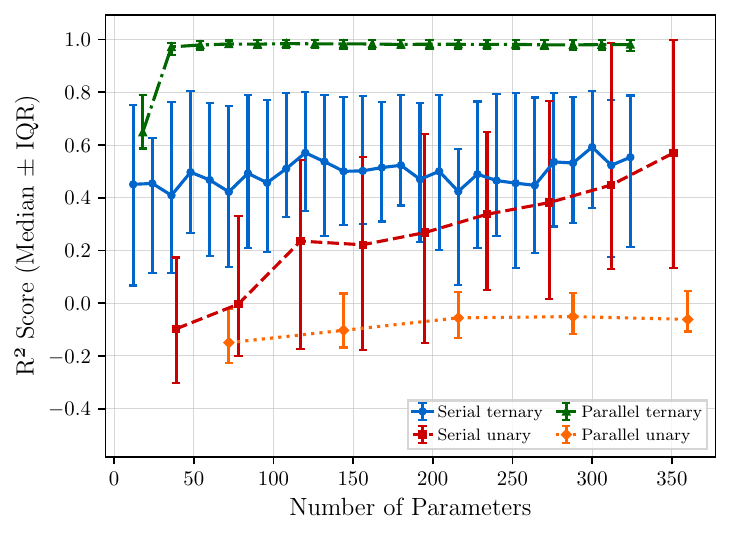}
    \caption{$R^2$ scores and IQRs for increasing numbers of model 
    parameters demonstrate how circuit architectures influence fitting 
    quality for the 1D partial Fourier series target function.}
    \label{fig:R2_Score_comparison_1dFourier_standard}
\end{figure}

The results in \cref{fig:R2_Score_comparison_1dFourier_standard} 
largely align with the theoretical requirements from 
\cref{sec:param_requirements}. The parallel ternary architecture on 
3 qubits quickly satisfies both the frequency control condition 
(27 frequencies) and the overparameterization condition $(2^4-2=14)$, achieving 
near-perfect fits. The parallel unary architecture fails the overparameterization condition $(2^{13}-2=8190)$, and the serial ternary 
architecture lacks sufficient individually trainable parameters to 
control 27 frequencies independently.

\begin{figure*}[t]
    \centering
    \begin{subfigure}{0.48\textwidth}
        \centering
        \includegraphics[width=\linewidth]{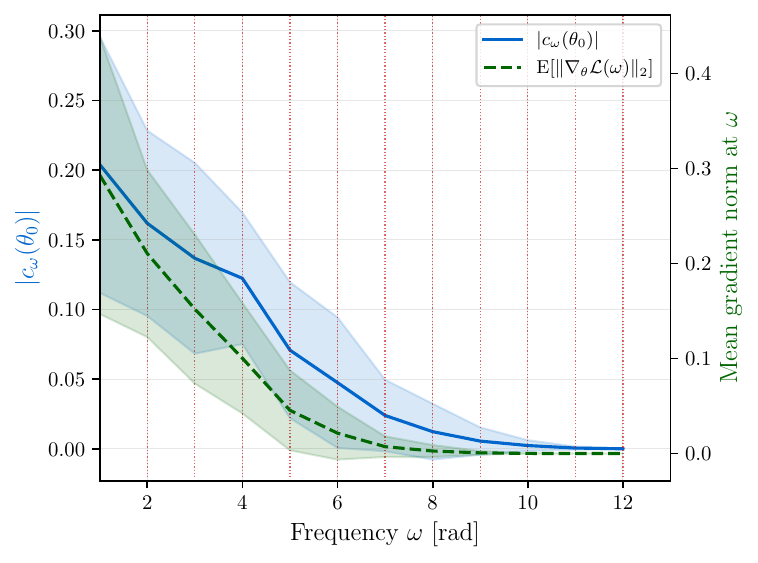}
        \caption{Fourier coefficient magnitudes $|c_\omega(\theta_0)|$ (blue) and 
        mean gradient norm $\mathrm{E}[\|\nabla_\theta \mathcal{L}(\omega)\|_2]$ 
        (green dashed) at initialization, averaged over 20 random seeds 
        ($\pm 1$ std shading). Both peak at low frequencies and decay toward 
        higher frequencies, tracking the redundancy profile $R(\omega)$.}
        \label{fig:spectral_gradient_init}
    \end{subfigure}
    \hfill
    \begin{subfigure}{0.48\textwidth}
        \centering
        \includegraphics[width=\linewidth]{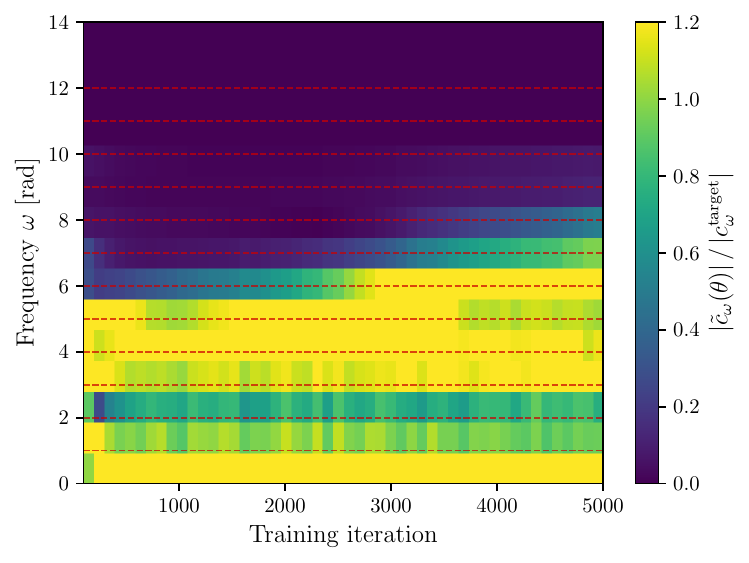}
        \caption{Frequency learning dynamics during training. 
        Low-frequency components are progressively fitted during training 
        while high-frequency targets ($\omega \geq 9$) 
        with low redundancy remain systematically near zero throughout, 
        consistent with redundancy competition as the failure mechanism.}
        \label{fig:freq_learning_heatmap}
    \end{subfigure}
    \caption{Spectral bias diagnostics for the serial unary architecture 
    (1 qubit, $L=12$, 39 parameters) on the dense 1D target 
    $\{1,\ldots,12\}$. Despite satisfying all three parameter 
    sufficiency conditions, the $\sum_\omega R(\omega) = 4^{12} \approx 
    16.8\mathrm{M}$ non-unique frequency instances dominate the gradient 
    landscape, crowding out target frequencies with low $R(\omega)$.}
    \label{fig:spectral_diagnostics}
\end{figure*}

The serial unary architecture yields a somewhat unexpected result. 
Despite satisfying all three conditions with only 39 parameters --- 
frequency coverage ($|\Omega| = 25$), frequency control, and the 
overparameterization condition ($2^{n+1}-2 = 2$) --- together with 
universal approximation guarantees, it consistently fails to fit the 
target function. This highlights an important distinction between 
\emph{universal approximation capability} and \emph{practical 
trainability}: the universal approximation result guarantees the 
existence of suitable parameters, but not that gradient-based 
optimization can find them. The serial unary spectrum contains 
$\sum_\omega R(\omega) = 4^{12} \approx 16.8\mathrm{M}$ total 
frequency instances for only 25 unique frequencies, and the resulting 
gradient dominance of high-redundancy non-unique frequencies crowds 
out the target frequencies regardless of parameter count. The spectral 
bias diagnostics in \cref{fig:spectral_diagnostics} confirm this: 
both the empirical Fourier spectrum and gradient norms at 
initialization follow the $R(\omega)$ profile, with high-frequency 
target components receiving disproportionately small gradient signals 
and remaining systematically unfitted throughout training.

To assess whether potential enhancements can mitigate the 
redundancy-driven spectral bias on the serial unary 1-qubit 
architecture, we compare Adam with standard initialization against 
four alternatives: small-angle initialization 
$\theta \sim \mathcal{N}(0, 1/L)$ \cite{Zhang_2025} combined with 
Adam, and the gradient-free optimizers CMA-ES, Nelder-Mead, and SPSA. 
As established in \cref{sec:initialization}, Zhang initialization 
reduces the redundancy-driven gradient dominance by suppressing all 
Fourier coefficients toward zero at the start of training. The 
gradient-free optimizers bypass the gradient magnitude bounds of 
\cite{Duffy_2026} entirely, allowing us to assess whether the fitting 
limitations are intrinsic to gradient-based optimization or to the 
spectral structure itself.

\begin{figure}[h]
    \centering
    \includegraphics[width=0.48\textwidth]{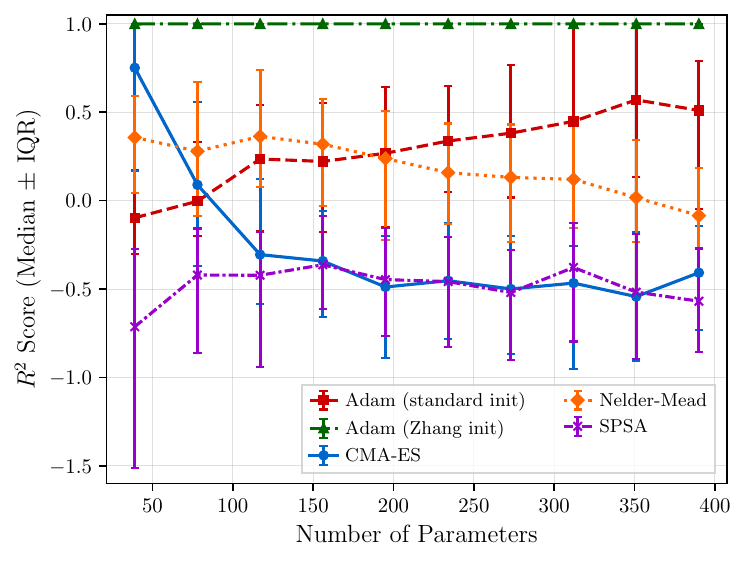}
    \caption{$R^2$ scores (median $\pm$ IQR) for the serial unary 
    encoding architecture on 1 qubit for the 1D target functions, 
    comparing Adam with standard initialization against four potential 
    enhancements: small-angle initialization $\theta \sim 
    \mathcal{N}(0, 1/L)$ \cite{Zhang_2025} combined with Adam, and 
    the gradient-free optimizers CMA-ES, Nelder-Mead, and SPSA. Zhang 
    initialization consistently achieves near-perfect $R^2 \approx 1$ 
    across all parameter counts, while all gradient-free optimizers 
    fail to match this performance.}
    \label{fig:R2_Score_comparison_1dFourier_zhang_gradfree}
\end{figure}

The results in \cref{fig:R2_Score_comparison_1dFourier_zhang_gradfree} 
show that Zhang initialization achieves near-perfect $R^2 \approx 1$ 
across all parameter counts, while all three gradient-free optimizers 
fail to match this performance, with median $R^2$ scores remaining 
well below 1 throughout. Taking the 1D results together, we identify 
the parallel ternary architecture combined with Zhang initialization 
as the most promising setup for the higher-dimensional experiments, 
as it achieves reliable convergence while benefiting from ternary 
encoding's reduced $R(\omega)$ profile.

\subsubsection{2D Synthetic Target Functions}
\label{sec:2d_param_requirements}

The 2D experiments examine how dense ternary encoding and frequency 
selection scale with input dimension.

\begin{figure}[htbp]
    \centering
    \includegraphics[width=0.48\textwidth]{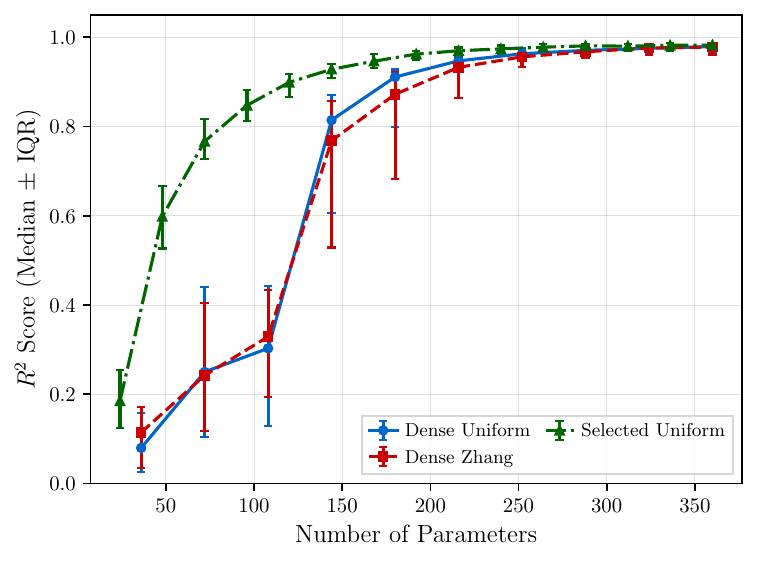}
    \caption{$R^2$ scores (median $\pm$ IQR) for the low-frequency 2D 
    target $\omega \in \{3,6,9,12\}^2$ on a 4-qubit parallel 
    architecture. Selected frequency encoding reaches $R^2 \approx 0.95$ 
    below 100 parameters, while dense ternary encoding under both 
    uniform and Zhang initialization requires more than twice as many 
    parameters to reach comparable performance, though all three 
    approaches converge to near-perfect fits at sufficient parameter 
    counts.}
    \label{fig:R2_Score_comparison_2dFourier_smallOmega}
\end{figure}

For the low-frequency 2D target with $\omega \in \{3,6,9,12\}^2$, 
shown in \cref{fig:R2_Score_comparison_2dFourier_smallOmega}, both 
dense initialization strategies---uniform and Zhang---approach 
near-perfect fits, but only at substantially higher parameter counts 
than the selected frequency approach. The selected frequency model 
achieves $R^2 \approx 0.95$ below 100 parameters, while the dense 
models require more than twice as many parameters to reach comparable 
performance. The two initialization strategies perform similarly once 
sufficient parameters are available.

\begin{figure}[htbp]
    \centering
    \includegraphics[width=0.48\textwidth]{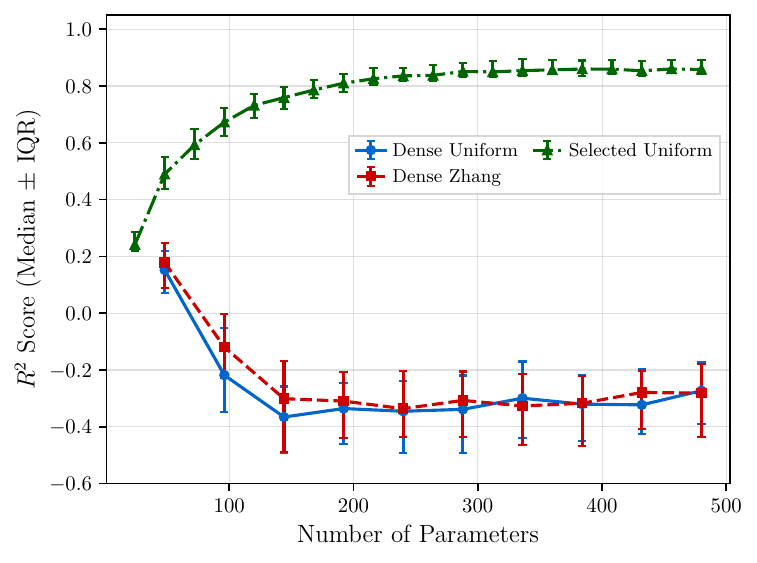}
    \caption{$R^2$ scores (median $\pm$ IQR) for the high-frequency 2D 
    target $\omega \in \{10,20,30,40\}^2$, comparing dense ternary 
    encoding under uniform and Zhang initialization against selected 
    frequency encoding. Dense approaches fail across the full parameter 
    range while frequency selection remains tractable.}
    \label{fig:R2_Score_comparison_2dFourier_largeOmega}
\end{figure}

The high-frequency 2D target with $\omega \in \{10,20,30,40\}^2$, 
shown in \cref{fig:R2_Score_comparison_2dFourier_largeOmega}, 
illustrates the resource intractability of dense encoding at scale. 
To cover frequencies up to 40, the dense baseline requires ternary 
prefactors $\{1,3,9,27\}$, yielding $81^2 = \num{6561}$ unique 
frequency pairs that the optimizer must jointly control --- 
far exceeding the 480-parameter budget and thus violating the 
frequency control condition ($p \geq |\Omega|$) by a factor of 
more than thirteen. Failure is therefore unsurprising from a parameter 
sufficiency standpoint alone. The more fundamental point is that satisfying condition 2 for 
this dense encoding would require $p \geq \num{6561}$ parameters, 
which demands a circuit depth that accumulates noise well beyond 
what current hardware can sustain coherently. By contrast, the selected frequency 
model with prefactors $\{10,30\}$ requires only $9^2 = 81$ unique 
frequency pairs --- an 81-fold reduction that restores parameter 
feasibility within the same budget. The selected frequency model 
improves steadily and achieves $R^2 \approx 0.85$, demonstrating 
that frequency selection remains tractable precisely where the dense 
parameter requirement becomes prohibitive.

\subsubsection{Real-World Dataset: CCPP}
\label{sec:ccpp_results}

On the CCPP dataset with the hardware-native ladder ansatz (96 
parameters, tbl$=2$), selected frequency encoding achieves a median 
$R^2$ of $0.923$ compared to $0.914$ for full ternary encoding, 
with substantially lower variance across ten runs (std $0.007$ vs 
$0.020$) and a worst-case $R^2$ of $0.901$ compared to $0.866$ for 
full ternary. Under depolarizing noise ($p=0.005$ per gate), both 
models perform comparably in terms of median $R^2$ ($0.907$ vs 
$0.909$), but the variance advantage of selected frequency encoding 
is markedly stronger (std $0.009$ vs $0.030$), with a worst-case 
$R^2$ of $0.896$ compared to $0.822$ for full ternary. The 
consistently lower variance across both clean and noisy conditions 
suggests a more stable optimization landscape, consistent with the 
reduced optimization burden from restricting $|\Omega|$ to only the 
frequencies identified in the per-feature spectral analysis.

\section{Discussion}
\label{sec:discussion}

\textbf{Interpretation and Related Work:}
Our white-box experiments systematically demonstrate the fundamental 
limitations of current quantum models with angle encoding when deployed 
with extensive frequency spectra, where required parameter counts 
rapidly exceed hardware capabilities. The spectral bias induced by 
frequency redundancy, as formalized by Duffy and 
Jastrzebski~\cite{Duffy_2026}, provides a theoretical explanation for 
these failures that goes beyond the three conditions established by 
prior work: it is not merely that parameters are insufficient in 
number, but that the gradient landscape is dominated by 
high-redundancy non-unique frequencies that crowd out the target 
frequencies regardless of circuit depth. This connects to the broader argument of Belis et al.\ 
\cite{Belis_2026} that spectral methods are fundamental to quantum 
machine learning: the redundancy structure acts as a double-edged 
sword, providing natural regularization through suppressed 
high-frequency gradients while simultaneously preventing the 
optimizer from fitting intended target frequencies.

Small angle weight initialization, as formalized by Zhang et 
al.\ \cite{Zhang_2025} and theoretically grounded in the redundancy 
framework of \cite{Duffy_2026}, offers a practical mitigation 
technique in one-dimensional settings by suppressing all Fourier 
coefficients initially toward zero and thereby reducing the gradient dominance 
of high-redundancy frequencies. The finding that gradient-free 
optimizers --- which bypass the gradient magnitude bounds of 
\cite{Duffy_2026} entirely --- fail to match this performance 
suggests that the benefit of Zhang initialization is not simply 
due to escaping a poorly conditioned gradient landscape, but 
reflects a more fundamental interaction between initialization 
scale and the redundancy structure of the model. However, as the high-frequency 2D experiments demonstrate, this 
represents a more fundamental resource intractability rather than 
an initialization problem: the dense encoding requires 
$p \geq \num{6561}$ parameters to satisfy condition 2, which 
exceeds the parameter budget available within a fixed circuit 
architecture. In our experiments, neither uniform nor Zhang 
initialization can compensate for this violated parameter 
sufficiency condition.

Our selected frequency approach addresses this at its source 
through strategic frequency reduction rather than exhaustive dense 
frequency space exploration, directly restoring parameter 
feasibility within the same budget.  As alternatives to address parameter 
scalability, Jaderberg et al.\ \cite{Jaderberg_2024} suggested 
training frequency prefactors as learnable parameters. However, 
as Yu et al.\ \cite{yu2024nonasymptoticapproximationerrorbounds} note, multiplying input data by 
trainable parameters during the encoding stage introduces a hybrid 
classical-quantum structure, making it difficult to attribute the 
resulting expressive power to the quantum component rather than the 
classical preprocessing. Frequency selection avoids this ambiguity 
by fixing the encoding structure before training, ensuring that the 
model's spectral properties are entirely determined by the quantum 
circuit architecture. Wiedmann et al.\ 
\cite{wiedmann2024fourieranalysisvariationalquantum} noted that 
systematically constraining parameters to 0 can remove frequencies 
from the model spectrum---a mechanism that can eliminate unnecessary 
frequencies but may also remove essential ones. Frequency selection 
operates at the encoding stage before training begins, making the 
frequency choices explicit and deliberate rather than an implicit 
side effect of optimization. However, as Wiedmann et al.\ note, 
ansatz parameter constraints can silently suppress selected 
frequencies even when they are present in the encoding spectrum. 
In cases of unexpectedly low performance, post-hoc spectral 
analysis of the trained model could help diagnose whether intended 
frequencies are actually being expressed, though developing 
systematic diagnostic procedures for this purpose is left for 
future work.

\textbf{Limitations and Scope of Applicability:}
Our approach has important limitations. First, our synthetic 
experimental targets were partial Fourier series themselves, 
ensuring perfect representability; real-world datasets may be less 
amenable to Fourier approximation, limiting achievable accuracy. 
Second, we assume knowledge of target frequency composition. For 
practical applications where this is unavailable, dominant 
frequencies can be identified via 1D Fourier analysis on dataset 
slices per dimension \cite{wiedmann2024fourieranalysisvariationalquantum}, 
with prefactors then chosen to cover only those frequencies. Our 
preliminary validation on the CCPP dataset, where prefactors were 
selected based on per-feature spectral analysis, suggests this 
strategy transfers to real-world settings, though further 
investigation is needed. In settings where even approximate 
spectral knowledge is unavailable, a conservative sparse spectrum 
could be iteratively expanded based on residual fitting errors. 
More broadly, even with frequency selection, mixed frequency 
growth remains a fundamental scaling challenge as input 
dimensionality increases.

\section{Conclusion}
\label{sec:conclusion}
Quantum machine learning faces exponential parameter scaling when 
representing target functions with dense Fourier spectra across 
multiple dimensions, creating parameter requirements that rapidly 
exceed current hardware capabilities. Building on the theoretical 
redundancy-gradient framework of Duffy and 
Jastrzebski~\cite{Duffy_2026}, this work provides systematic 
experimental evidence that the redundancy structure of the 
frequency spectrum is a key source of these failures in the unary 
encoding setting: non-unique frequencies with high redundancy 
$R(\omega)$ dominate the gradient landscape and crowd out the 
target frequencies. More broadly, even with ternary encoding where per-frequency 
redundancy is minimized, the combinatorial growth of unique 
frequency tuples with input dimension drives the parameter 
requirement $p \geq |\Omega|$ beyond any realistic budget --- 
a resource intractability that no initialization or optimizer 
choice can resolve without first reducing $|\Omega|$ itself.

Small-angle weight initialization in the form of the adaptive 
Gaussian scheme of Zhang et al.\ \cite{Zhang_2025} mitigates this 
burden in one-dimensional settings by suppressing the 
redundancy-driven gradient dominance, allowing the optimizer to 
reach target frequencies more effectively. Notably, gradient-free optimizers that bypass the gradient 
magnitude bounds entirely also fail to achieve competitive 
performance, indicating that the fitting limitations on this 
architecture are not solely an artifact of gradient-based 
optimization. However, our two-dimensional experiments confirm that 
initialization-based mitigation reaches a fundamental ceiling once 
the parameter requirement $p \geq |\Omega|$ exceeds the available 
budget: in our experiments neither initialization strategy can 
compensate for this violated condition, and no amount of additional 
parameters rescues the dense model within a fixed circuit architecture.

Frequency selection addresses this limitation at its source by 
restricting the model spectrum to only those frequencies present 
in the target function, reducing $|\Omega|$ and with it the 
parameter requirement, keeping circuit depth tractable even as 
target frequency magnitudes grow. Our results on both synthetic and real-world data show that this 
approach enables more effective use of the available parameter 
budget, achieving competitive performance where dense frequency 
strategies fail, and remains tractable at frequency magnitudes 
where dense approaches become entirely infeasible. Under current 
hardware constraints, matching the model's frequency spectrum to 
the task at hand --- rather than maximizing spectral coverage --- 
offers a principled and practical path toward quantum machine 
learning models that respect both hardware limitations and the 
fundamental trainability constraints identified here.

\section*{Acknowledgment}
This work has been supported by the LMU Sustainability Fund (EfOiE), the BMFTR (QuCUN, QuaRDS, CAQAO), the Munich Quantum Valley (K5, K7), and the Bavarian StMWi (6GQT). The sole responsibility for the report's contents lies with the authors. 
Claude (Anthropic) was used to assist with writing and editing 
throughout all sections, to debug and generate experimental and plotting code 
(\cref{sec:results,sec:exp_setup_details}), in accordance with IEEE policy 
on AI-generated content. All AI-assisted content was reviewed and 
verified by the authors.

\bibliographystyle{IEEEtran}
\bibliography{icml2026}

@article{Ragone_2024,
   title={A Lie algebraic theory of barren plateaus for deep parameterized quantum circuits},
   volume={15},
   ISSN={2041-1723},
   url={http://dx.doi.org/10.1038/s41467-024-49909-3},
   DOI={10.1038/s41467-024-49909-3},
   number={1},
   journal={Nature Communications},
   publisher={Springer Science and Business Media LLC},
   author={Ragone, Michael and Bakalov, Bojko N. and Sauvage, Frédéric and Kemper, Alexander F. and Ortiz Marrero, Carlos and Larocca, Martín and Cerezo, M.},
   year={2024},
   month=aug }

@article{Schuld_2021,
   title={Effect of data encoding on the expressive power of variational quantum-machine-learning models},
   volume={103},
   ISSN={2469-9934},
   url={http://dx.doi.org/10.1103/PhysRevA.103.032430},
   DOI={10.1103/physreva.103.032430},
   number={3},
   journal={Physical Review A},
   publisher={American Physical Society (APS)},
   author={Schuld, Maria and Sweke, Ryan and Meyer, Johannes Jakob},
   year={2021},
   month=mar }

@article{Jaderberg_2024,
   title={Let quantum neural networks choose their own frequencies},
   volume={109},
   ISSN={2469-9934},
   url={http://dx.doi.org/10.1103/PhysRevA.109.042421},
   DOI={10.1103/physreva.109.042421},
   number={4},
   journal={Physical Review A},
   publisher={American Physical Society (APS)},
   author={Jaderberg, Ben and Gentile, Antonio A. and Berrada, Youssef Achari and Shishenina, Elvira and Elfving, Vincent E.},
   year={2024},
   month=apr }

@article{Kordzanganeh_2023,
   title={An exponentially-growing family of universal quantum circuits},
   volume={4},
   ISSN={2632-2153},
   url={http://dx.doi.org/10.1088/2632-2153/ace757},
   DOI={10.1088/2632-2153/ace757},
   number={3},
   journal={Machine Learning: Science and Technology},
   publisher={IOP Publishing},
   author={Kordzanganeh, Mo and Sekatski, Pavel and Fedichkin, Leonid and Melnikov, Alexey},
   year={2023},
   month=aug, pages={035036} }

@article{Shin_2023,
   title={Exponential data encoding for quantum supervised learning},
   volume={107},
   ISSN={2469-9934},
   url={http://dx.doi.org/10.1103/PhysRevA.107.012422},
   DOI={10.1103/physreva.107.012422},
   number={1},
   journal={Physical Review A},
   publisher={American Physical Society (APS)},
   author={Shin, S. and Teo, Y. S. and Jeong, H.},
   year={2023},
   month=jan }

@article{Peters_2023,
   title={Generalization despite overfitting in quantum machine learning models},
   volume={7},
   ISSN={2521-327X},
   url={http://dx.doi.org/10.22331/q-2023-12-20-1210},
   DOI={10.22331/q-2023-12-20-1210},
   journal={Quantum},
   publisher={Verein zur Forderung des Open Access Publizierens in den Quantenwissenschaften},
   author={Peters, Evan and Schuld, Maria},
   year={2023},
   month=dec, pages={1210} }

@article{Larocca_2023,
   title={Theory of overparametrization in quantum neural networks},
   volume={3},
   ISSN={2662-8457},
   url={http://dx.doi.org/10.1038/s43588-023-00467-6},
   DOI={10.1038/s43588-023-00467-6},
   number={6},
   journal={Nature Computational Science},
   publisher={Springer Science and Business Media LLC},
   author={Larocca, Martín and Ju, Nathan and García-Martín, Diego and Coles, Patrick J. and Cerezo, Marco},
   year={2023},
   month=jun, pages={542–551} }

@article{Mitarai_2018,
   title={Quantum circuit learning},
   volume={98},
   ISSN={2469-9934},
   url={http://dx.doi.org/10.1103/PhysRevA.98.032309},
   DOI={10.1103/physreva.98.032309},
   number={3},
   journal={Physical Review A},
   publisher={American Physical Society (APS)},
   author={Mitarai, K. and Negoro, M. and Kitagawa, M. and Fujii, K.},
   year={2018},
   month=sep }

@misc{holzer2024spectralinvariancemaximalityproperties,
      title={Spectral invariance and maximality properties of the frequency spectrum of quantum neural networks}, 
      author={Patrick Holzer and Ivica Turkalj},
      year={2024},
      eprint={2402.14515},
      archivePrefix={arXiv},
      primaryClass={quant-ph},
      url={https://arxiv.org/abs/2402.14515}, 
}

@article{Abbas_2021,
   title={The power of quantum neural networks},
   volume={1},
   ISSN={2662-8457},
   url={http://dx.doi.org/10.1038/s43588-021-00084-1},
   DOI={10.1038/s43588-021-00084-1},
   number={6},
   journal={Nature Computational Science},
   publisher={Springer Science and Business Media LLC},
   author={Abbas, Amira and Sutter, David and Zoufal, Christa and Lucchi, Aurelien and Figalli, Alessio and Woerner, Stefan},
   year={2021},
   month=jun, pages={403–409} }

@book{nielsen2002quantum,
  title={Quantum Computation and Quantum Information},
  author={Nielsen, Michael A and Chuang, Isaac L},
  year={2002},
  publisher={Cambridge University Press}
}

@misc{optax2020github,
  author       = {DeepMind},
  title        = {Optax: Gradient processing and optimization library in JAX},
  year         = {2020},
  howpublished = {\url{https://optax.readthedocs.io/en/latest/api/optimizers.html}},
  note         = {Accessed: 2025-07-29}
}

@Article{ijms26136325,
AUTHOR = {Niazi, Sarfaraz K.},
TITLE = {Quantum Mechanics in Drug Discovery: A Comprehensive Review of Methods, Applications, and Future Directions},
JOURNAL = {International Journal of Molecular Sciences},
VOLUME = {26},
YEAR = {2025},
NUMBER = {13},
ARTICLE-NUMBER = {6325},
URL = {https://www.mdpi.com/1422-0067/26/13/6325},
PubMedID = {40650102},
ISSN = {1422-0067},
DOI = {10.3390/ijms26136325}
}

@booklet{EasyChair:15051,
  author    = {Wayzman Kolawole},
  title     = {Hybrid Financial Models for Capturing Asset Correlation Structures},
  howpublished = {EasyChair Preprint 15051},
  year      = {2024}
}

@misc{bergholm2022pennylaneautomaticdifferentiationhybrid,
      title={PennyLane: Automatic differentiation of hybrid quantum-classical computations}, 
      author={Ville Bergholm and Josh Izaac and Maria Schuld and Christian Gogolin and Shahnawaz Ahmed and Vishnu Ajith and M. Sohaib Alam and Guillermo Alonso-Linaje and B. AkashNarayanan and Ali Asadi and Juan Miguel Arrazola and Utkarsh Azad and Sam Banning and Carsten Blank and Thomas R Bromley and Benjamin A. Cordier and Jack Ceroni and Alain Delgado and Olivia Di Matteo and Amintor Dusko and Tanya Garg and Diego Guala and Anthony Hayes and Ryan Hill and Aroosa Ijaz and Theodor Isacsson and David Ittah and Soran Jahangiri and Prateek Jain and Edward Jiang and Ankit Khandelwal and Korbinian Kottmann and Robert A. Lang and Christina Lee and Thomas Loke and Angus Lowe and Keri McKiernan and Johannes Jakob Meyer and J. A. Montañez-Barrera and Romain Moyard and Zeyue Niu and Lee James O'Riordan and Steven Oud and Ashish Panigrahi and Chae-Yeun Park and Daniel Polatajko and Nicolás Quesada and Chase Roberts and Nahum Sá and Isidor Schoch and Borun Shi and Shuli Shu and Sukin Sim and Arshpreet Singh and Ingrid Strandberg and Jay Soni and Antal Száva and Slimane Thabet and Rodrigo A. Vargas-Hernández and Trevor Vincent and Nicola Vitucci and Maurice Weber and David Wierichs and Roeland Wiersema and Moritz Willmann and Vincent Wong and Shaoming Zhang and Nathan Killoran},
      year={2022},
      eprint={1811.04968},
      archivePrefix={arXiv},
      primaryClass={quant-ph},
      url={https://arxiv.org/abs/1811.04968}, 
}

@article{Kandala_2017,
   title={Hardware-efficient variational quantum eigensolver for small molecules and quantum magnets},
   volume={549},
   ISSN={1476-4687},
   url={http://dx.doi.org/10.1038/nature23879},
   DOI={10.1038/nature23879},
   number={7671},
   journal={Nature},
   publisher={Springer Science and Business Media LLC},
   author={Kandala, Abhinav and Mezzacapo, Antonio and Temme, Kristan and Takita, Maika and Brink, Markus and Chow, Jerry M. and Gambetta, Jay M.},
   year={2017},
   month=sep, pages={242–246} }

@article{Wiersema_2024,
   title={Classification of dynamical Lie algebras of 2-local spin systems on linear, circular and fully connected topologies},
   volume={10},
   ISSN={2056-6387},
   url={http://dx.doi.org/10.1038/s41534-024-00900-2},
   DOI={10.1038/s41534-024-00900-2},
   number={1},
   journal={npj Quantum Information},
   publisher={Springer Science and Business Media LLC},
   author={Wiersema, Roeland and Kökcü, Efekan and Kemper, Alexander F. and Bakalov, Bojko N.},
   year={2024},
   month=nov }

@article{Cerezo_2021,
   title={Variational quantum algorithms},
   volume={3},
   ISSN={2522-5820},
   url={http://dx.doi.org/10.1038/s42254-021-00348-9},
   DOI={10.1038/s42254-021-00348-9},
   number={9},
   journal={Nature Reviews Physics},
   publisher={Springer Science and Business Media LLC},
   author={Cerezo, M. and Arrasmith, Andrew and Babbush, Ryan and Benjamin, Simon C. and Endo, Suguru and Fujii, Keisuke and McClean, Jarrod R. and Mitarai, Kosuke and Yuan, Xiao and Cincio, Lukasz and Coles, Patrick J.},
   year={2021},
   month=aug, pages={625–644} }

@article{pedregosa2011scikit,
  title={Scikit-learn: Machine learning in Python},
  author={Pedregosa, Fabian and Varoquaux, Ga{\"e}l and Gramfort, Alexandre and Michel, Vincent and Thirion, Bertrand and Grisel, Olivier and Blondel, Mathieu and Prettenhofer, Peter and Weiss, Ron and Dubourg, Vincent and others},
  journal={Journal of machine learning research},
  volume={12},
  number={Oct},
  pages={2825--2830},
  year={2011}
}

@misc{wiedmann2024fourieranalysisvariationalquantum,
      title={Fourier Analysis of Variational Quantum Circuits for Supervised Learning}, 
      author={Marco Wiedmann and Maniraman Periyasamy and Daniel D. Scherer},
      year={2024},
      eprint={2411.03450},
      archivePrefix={arXiv},
      primaryClass={cs.LG},
      url={https://arxiv.org/abs/2411.03450}, 
}

@misc{farhi2018classificationquantumneuralnetworks,
      title={Classification with Quantum Neural Networks on Near Term Processors}, 
      author={Edward Farhi and Hartmut Neven},
      year={2018},
      eprint={1802.06002},
      archivePrefix={arXiv},
      primaryClass={quant-ph},
      url={https://arxiv.org/abs/1802.06002}, 
}

@misc{yu2024nonasymptoticapproximationerrorbounds,
      title={Non-asymptotic Approximation Error Bounds of Parameterized Quantum Circuits}, 
      author={Zhan Yu and Qiuhao Chen and Yuling Jiao and Yinan Li and Xiliang Lu and Xin Wang and Jerry Zhijian Yang},
      year={2024},
      eprint={2310.07528},
      archivePrefix={arXiv},
      primaryClass={quant-ph},
      url={https://arxiv.org/abs/2310.07528}, 
}

@misc{Belis_2026,
      title={Spectral methods: crucial for machine learning, natural for quantum computers?}, 
      author={Vasilis Belis and Joseph Bowles and Rishabh Gupta and Evan Peters and Maria Schuld},
      year={2026},
      eprint={2603.24654},
      archivePrefix={arXiv},
      primaryClass={quant-ph},
      url={https://arxiv.org/abs/2603.24654}, 
}

@misc{Rahaman_2019,
      title={On the Spectral Bias of Neural Networks}, 
      author={Nasim Rahaman and Aristide Baratin and Devansh Arpit and Felix Draxler and Min Lin and Fred A. Hamprecht and Yoshua Bengio and Aaron Courville},
      year={2019},
      eprint={1806.08734},
      archivePrefix={arXiv},
      primaryClass={stat.ML},
      url={https://arxiv.org/abs/1806.08734}, 
}

@article{Zhang_2025,
  author  = {Zhang, Kaining and Liu, Liu and Hsieh, Min-Hsiu and Tao, Dacheng},
  title   = {Escaping from the Barren Plateau via {G}aussian Initializations 
             in Deep Variational Quantum Circuits},
  journal = {Physical Review Letters},
  year    = {2025},
  volume  = {134},
  pages   = {150601},
  doi     = {10.1103/PhysRevLett.134.150601}
}

@article{Duffy_2026,
  author  = {Duffy, Callum and Jastrzebski, Marcin},
  title   = {Spectral Bias in Variational Quantum Machine Learning},
  journal = {arXiv preprint arXiv:2506.22555},
  year    = {2026},
  note    = {v2, January 2026}
}

@article{Perez_Salinas_2021,
   title={One qubit as a universal approximant},
   volume={104},
   ISSN={2469-9934},
   url={http://dx.doi.org/10.1103/PhysRevA.104.012405},
   DOI={10.1103/physreva.104.012405},
   number={1},
   journal={Physical Review A},
   publisher={American Physical Society (APS)},
   author={Pérez-Salinas, Adrián and López-Núñez, David and García-Sáez, Artur and Forn-Díaz, P. and Latorre, José I.},
   year={2021},
   month=jul }

@misc{combined_cycle_power_plant_294,
  author       = {Tfekci, Pnar and Kaya, Heysem},
  title        = {{Combined Cycle Power Plant}},
  year         = {2014},
  howpublished = {UCI Machine Learning Repository},
  note         = {{DOI}: https://doi.org/10.24432/C5002N}
}

@article{nelder1965simplex,
    author = {Nelder, J. A. and Mead, R.},
    title = {A Simplex Method for Function Minimization},
    journal = {The Computer Journal},
    volume = {7},
    number = {4},
    pages = {308-313},
    year = {1965},
    month = {01},
    issn = {0010-4620},
    doi = {10.1093/comjnl/7.4.308},
    url = {https://doi.org/10.1093/comjnl/7.4.308},
    eprint = {https://academic.oup.com/comjnl/article-pdf/7/4/308/1013182/7-4-308.pdf},
}

@article{spall1998spsa,
  title={AN OVERVIEW OF THE SIMULTANEOUS PERTURBATION METHOD FOR EFFICIENT OPTIMIZATION},
  author={James C. Spall},
  journal={Johns Hopkins Apl Technical Digest},
  year={1998},
  volume={19},
  pages={482-492},
  url={https://api.semanticscholar.org/CorpusID:7988308}
}

@article{hansen2016cma,
  title={The CMA Evolution Strategy: A Tutorial},
  author={Nikolaus Hansen},
  journal={ArXiv},
  year={2016},
  volume={abs/1604.00772},
  url={https://api.semanticscholar.org/CorpusID:15038271}
}

@software{jax2018github,
  author  = {James Bradbury and Roy Frostig and Peter Hawkins and
             Matthew James Johnson and Chris Leary and Dougal Maclaurin and
             George Necula and Adam Paszke and Jake Vander{P}las and
             Skye Wanderman-{M}ilne and Qiao Zhang},
  title   = {{JAX}: composable transformations of {P}ython+{N}um{P}y programs},
  url     = {http://github.com/jax-ml/jax},
  version = {0.9.0.1},
  year    = {2018},
}

@article{ibm_heron_r2_2025,
  title   = {Characterising the failure mechanisms of error-corrected quantum logic gates},
  author  = {{IBM Quantum} and others},
  journal = {arXiv preprint arXiv:2504.07258},
  year    = {2025},
  eprint  = {2504.07258},
  archivePrefix = {arXiv},
  primaryClass  = {quant-ph}
}

\clearpage
\appendices

\section{Experimental Setup Details}
\label{sec:exp_setup_details}

\subsection{Target Function Design}

For the one-dimensional case, we define the target function as:
\begin{equation}
t_{1}(x) = c_0 + \sum_{i=1}^4 \left[ a_i \cos(\omega_ix) + 
b_i\sin(\omega_ix) \right]
\label{eq:target_1d}
\end{equation}
where $\omega_i \in \{3, 6, 9, 12\}$. The coefficients $c_0, a_i, b_i$ 
are randomly sampled from $[0,1)$ to generate 10 distinct instances 
with identical spectral characteristics but varying amplitudes.

The two-dimensional target function extends this to include mixed 
frequency terms:
\begin{equation}
\begin{aligned}
t_2(x,y) = c_0 + \sum_{i=1}^{4} \sum_{j=1}^{4} &\bigg[ 
a_{ij} \cos(\omega_i^x x) \cos(\omega_j^y y) \\
&\quad + b_{ij} \cos(\omega_i^x x) \sin(\omega_j^y y) \\
&\quad + c_{ij} \sin(\omega_i^x x) \cos(\omega_j^y y) \\
&\quad + d_{ij} \sin(\omega_i^x x) \sin(\omega_j^y y) \bigg]
\end{aligned}
\label{eq:target_2d}
\end{equation}
with low-frequency experiments using $\omega_i^x, \omega_j^y \in 
\{3,6,9,12\}$ and high-frequency experiments using $\omega_i^x, 
\omega_j^y \in \{10,20,30,40\}$. The latter increases the total number of frequency instances in 
the model spectrum from $64^2 = \num{4096}$ to 
$256^2 = \num{65536}$, providing a more demanding test of the 
parameter and optimization burden as dimensionality and target 
frequency magnitude increase.

\subsection{Data Preparation and Implementation}

We use evenly spaced points across $[-\pi, \pi]$ for each input 
dimension (50 points for 1D; $25{\times}25$ grid for 2D). Inputs are 
scaled to $[-\pi, \pi]$ and targets to $[-1,1]$ via 
\texttt{MinMaxScaler}~\cite{pedregosa2011scikit}. All experiments use 
an 80/20 train-test split and are implemented in PennyLane 
(v0.42.0)~\cite{bergholm2022pennylaneautomaticdifferentiationhybrid} 
with JAX/JIT (v0.5.0)~\cite{jax2018github}, executed on a MacBook Pro 
with Apple M2, 16\,GB RAM, macOS Sequoia 15.5.

The circuit ansatz uses \texttt{qml.Rot}$(\boldsymbol{\theta})$ gates 
decomposed as $R_z(\theta_1)R_y(\theta_2)R_z(\theta_3)$, combined with 
CNOT gates in a ladder structure. Rotational gates are applied only 
after receiving the target element of a CNOT gate from the neighboring 
qubit above, ensuring all trainable parameters lie within the 
measurement lightcone of the observable~\cite{Cerezo_2021}.

\subsection{Training Configuration}

Unless otherwise stated, all models are trained with the Adam 
optimizer~\cite{optax2020github} at learning rate $\eta{=}0.001$ for 
$5{,}000$ steps, MSE loss, and evaluated by $R^2$ score. For the 
synthetic experiments we execute 10 independent runs across each of the 
10 target function realizations (100 experiments per configuration); 
results are summarized by median $R^2$ with IQR error bars.

\subsection{Benchmarking against Gradient-Free optimizers}

To verify that the failure of the 1-qubit model is attributable to 
spectral limitations rather than optimizer choice, we train a serial 
unary QNN ($n{=}1$, 12 encoding layers, $\alpha{=}1$ throughout) on 
the same 10 synthetic target functions, comparing Adam, 
SPSA~\cite{spall1998spsa}, CMA-ES~\cite{hansen2016cma}, and 
Nelder-Mead~\cite{nelder1965simplex} across depths $d\in\{1,\ldots,10\}$ 
trainable block layers. Adam and SPSA run for $5{,}000$ steps (batch 
size 40); CMA-ES for $2{,}000$ generations ($\sigma_0{=}0.5$); 
Nelder-Mead up to $50{,}000$ function evaluations. Each configuration 
uses 10 independent random initializations per target function.

\subsection{Real-World Dataset: Combined Cycle Power Plant}
\label{sec:ccpp_setup}

We evaluate frequency selection on the CCPP 
dataset~\cite{combined_cycle_power_plant_294} ($9{,}568$ samples; 
features: ambient temperature AT, exhaust vacuum V, ambient pressure 
AP, relative humidity RH; target: net electrical output PE). We 
construct a four-qubit parallel-encoding model with one qubit per 
feature and $\mathrm{RX}$ encoding gates, using the hardware-native 
ladder ansatz with 2 trainable block layers (96 parameters). We 
compare \emph{full ternary} (prefactors $1,3,9$ on all qubits) 
against \emph{selected frequency} (prefactors $1,3$ for AT/V/RH; 
$1,3,10$ for AP, identified via per-feature 1D Fourier analysis), 
trained with Adam ($3{,}000$ steps, batch size $64$, Zhang 
initialization \cite{Zhang_2025}) over 10 runs.
To additionally assess robustness under realistic hardware noise, 
both models are trained and evaluated under depolarizing 
noise~\cite{nielsen2002quantum} ($p{=}0.005$ per gate) via 
\texttt{default.mixed}~\cite{bergholm2022pennylaneautomaticdifferentiationhybrid} 
(batch size $32$, $3{,}000$ steps). Depolarizing channels are 
applied after every individual gate --- each single-qubit Rot and 
RX gate, and both qubits of each CNOT --- consistent with per-gate 
error rates on current superconducting 
hardware~\cite{ibm_heron_r2_2025}. This rate is applied uniformly 
to single-qubit gates as well, which in practice exhibit 
approximately an order of magnitude lower error rates, making the 
model a conservative noise estimate overall. The noise model 
accounts for gate-level depolarizing errors only; readout errors, 
decoherence, and crosstalk are not modeled, so the simulated noise 
level represents a lower bound on total hardware noise.

\clearpage
\onecolumn
\section{Circuit Architectures}
\label{sec:circuit_architectures}

We present three representative circuit architectures used in our experiments. 
Full circuit diagrams for all variants are available in the extended version of 
this paper on arXiv.

\begin{figure}[h]
\centering
\scalebox{0.9}{
\begin{quantikz}
    \lstick{$\ket{0}$} & \qw\gategroup[1,steps=3,style={dashed,rounded corners, inner xsep=2pt, inner ysep=10pt, yshift=-4pt},background,label style={label position=above,anchor=north,yshift=0.4cm}]{{\sc Ansatz 0}} & \gate{R(\boldsymbol{\theta_0})} & \qw  & \gate{Rx(x)}\gategroup[1,steps=1,style={dashed,rounded corners, inner xsep=2pt, inner ysep=10pt, yshift=-4pt},background,label style={label position=above,anchor=north,yshift=0.4cm}]{{\sc FM1}} & \qw\gategroup[1,steps=3,style={dashed,rounded corners, inner xsep=2pt, inner ysep=10pt, yshift=-4pt},background,label style={label position=above,anchor=north,yshift=0.4cm}]{{\sc Ansatz 1}} & \gate{R(\boldsymbol{\theta_1})}  & \qw & \ldots & \gate{Rx(x)}\gategroup[1,steps=1,style={dashed,rounded corners, inner xsep=2pt, inner ysep=10pt, yshift=-4pt},background,label style={label position=above,anchor=north,yshift=0.4cm}]{{\sc FM12}} & \qw\gategroup[1,steps=3,style={dashed,rounded corners, inner xsep=2pt, inner ysep=10pt, yshift=-4pt},background,label style={label position=above,anchor=north,yshift=0.4cm}]{{\sc Ansatz 12}}  & \gate{R(\boldsymbol{\theta_{12}})} & \qw & \meter{}
\end{quantikz}}
\caption{Serial unary circuit architecture for 1D target. To increase the number of parameters, additional general rotational gates $R(\boldsymbol{\theta})$ are added by including additional ansatz layer blocks between the FMs.}
\label{fig:serial_unary_circuit_1D}
\end{figure}

\begin{figure}[h]
\centering
\scalebox{0.8}{
\begin{quantikz}
    \lstick{$\ket{0}$} & \gate{R(\boldsymbol{\theta_0})}\gategroup[3,steps=7,style={dashed,rounded corners, inner xsep=2pt, inner ysep=10pt, yshift=-5pt},background,label style={label position=above,anchor=north,yshift=0.4cm}]{{\sc Ansatz 0}} &\ctrl{1} & \qw  & \qw & \qw & \qw & \qw & \gate{Rx(x)}\gategroup[3,steps=1,style={dashed,rounded corners, inner xsep=2pt, inner ysep=10pt, yshift=-5pt},background,label style={label position=above,anchor=north,yshift=0.4cm}]{{\sc FM1}} & \gate{R(\boldsymbol{\theta_{3}})}\gategroup[3,steps=7,style={dashed,rounded corners, inner xsep=2pt, inner ysep=10pt, yshift=-5pt},background,label style={label position=above,anchor=north,yshift=0.4cm}]{{\sc Ansatz 1}} &\ctrl{1}  & \qw & \qw & \qw & \qw & \qw & \qw\\
    \lstick{$\ket{0}$} & \qw & \targ{} & \gate{R(\boldsymbol{\theta_{1}})} &\ctrl{1} & \qw & \qw & \qw   & \gate{Rx(3x)} & \qw & \targ{} & \gate{R(\boldsymbol{\theta_{4}})} & \ctrl{1} & \qw & \qw & \qw & \qw \\
    \lstick{$\ket{0}$} & \qw & \qw & \qw & \targ{} & \gate{R(\boldsymbol{\theta_{2}})} &\qw & \qw  & \gate{Rx(9x)} & \qw & \qw & \qw & \targ{} & \gate{R(\boldsymbol{\theta_{5}})} &\qw  & \qw   & \meter{}
\end{quantikz}
}
\caption{Parallel ternary circuit architecture for 1D target. Within each ansatz layer, there is a CNOT connection to the following qubit after the rotational gate. FMs contain prefactors of $3^{l-1}$ for each vertical layer $l=1,\ldots,3$. To increase the number of parameters, additional general rotational gates $R(\boldsymbol{\theta})$ are added by including additional ansatz layer blocks between the FMs.}
\label{fig:parallel_ternary_circuit_1D}
\end{figure}

\begin{figure}[h]
\centering
\scalebox{0.8}{
\begin{quantikz}
    \lstick{$\ket{0}$} & \gate{R(\boldsymbol{\theta_0})}\gategroup[4,steps=7,style={dashed,rounded corners, inner xsep=2pt, inner ysep=10pt, yshift=-5pt},background,label style={label position=above,anchor=north,yshift=0.4cm}]{{\sc Ansatz 0}} &\ctrl{1}   & \qw & \qw & \qw & \qw & \qw  & \gate{Rx(3x)}\gategroup[4,steps=1,style={dashed,rounded corners, inner xsep=2pt, inner ysep=10pt, yshift=-5pt},background,label style={label position=above,anchor=north,yshift=0.4cm}]{{\sc FM1}} & \gate{R(\boldsymbol{\theta_{4}})}\gategroup[4,steps=7,style={dashed,rounded corners, inner xsep=2pt, inner ysep=10pt, yshift=-5pt},background,label style={label position=above,anchor=north,yshift=0.4cm}]{{\sc Ansatz 1}} &\ctrl{1}   & \qw & \qw & \qw & \qw & \qw & \qw \\
    \lstick{$\ket{0}$} & \qw & \targ{} & \gate{R(\boldsymbol{\theta_{1}})} &\ctrl{1} & \qw & \qw & \qw  & \gate{Rx(9x)} & \qw & \targ{} & \gate{R(\boldsymbol{\theta_{5}})} & \ctrl{1}  & \qw & \qw  & \qw & \qw \\
    \lstick{$\ket{0}$} & \qw & \qw & \qw & \targ{} & \gate{R(\boldsymbol{\theta_{2}})} & \ctrl{1} & \qw  & \gate{Rx(3y)} & \qw & \qw & \qw & \targ{} & \gate{R(\boldsymbol{\theta_{6}})} & \ctrl{1} & \qw  & \qw \\
    \lstick{$\ket{0}$} & \qw & \qw & \qw & \qw  & \qw & \targ{} & \gate{R(\boldsymbol{\theta_{3}})}   & \gate{Rx(9y)} & \qw & \qw  & \qw & \qw & \qw & \targ{} & \gate{R(\boldsymbol{\theta_{7}})}  & \meter{}
\end{quantikz}
}
\caption{Selected frequencies parallel circuit architecture (2D synthetic 
experiment, 4 qubits). Within each ansatz layer, there is a CNOT 
connection to the following qubit after the rotational gate. FMs 
contain prefactors of $3$ and $9$ for each consecutive FM in each 
dimension, covering the target frequencies $\{3,6,9,12\}^2$. To 
increase the number of parameters, additional general rotational 
gates $R(\boldsymbol{\theta})$ are added by including additional 
ansatz layer blocks between the FMs.}
\label{fig:sel_parallel_circuit_4Q_2D}
\end{figure}

\end{document}